\documentclass{appolb}

\RequirePackage[utf8]{inputenc}

\RequirePackage{color}
\RequirePackage{amssymb}
\RequirePackage{graphicx}

\DeclareMathAlphabet{\mathcal}{OMS}{cmsy}{m}{n}

\def\d{{\rm d}}
\def\un#1{\,{\rm #1}}

\def\e{{\rm e}}
\def\I{{\rm i}}

\def\NEW#1{{#1}}

\setbox123\hbox{$0$}
\setbox124\hbox{$.$}
\def\S{\hbox to\wd123{\hss}}
\def\.{\hbox to\wd124{\hss}}

\begin{document}

\title{Coulomb-nuclear interference in elastic scattering: eikonal calculation to all orders of $\alpha$}

\author{
	J.~Ka\v spar
		\address{
			Institute of Physics of the Academy of Sciences of the Czech Republic, Prague, Czech Republic
			and CERN, Geneva, Switzerland.
		}
}

\maketitle

\begin{abstract}
The Coulomb-nuclear interference (CNI) has recently been used by the TOTEM Collaboration to analyse proton-proton elastic-scattering data from the LHC and to draw physics conclusions. This paper will present an eikonal calculation of the CNI effects performed to all orders of the fine structure constant, $\alpha$. This calculation will be used as a reference to benchmark several widely-used CNI formulae and to verify several recent claims by other authors.
\end{abstract}


%----------------------------------------------------------------------------------------------------

%\linenumbers

%----------------------------------------------------------------------------------------------------

\section{Introduction}
\label{sec:introduction}

Elastic scattering of nucleons is a process mediated by electromagnetic (Coulomb) and strong (nuclear) force. In the domain of small squared four-momentum transfer, $|t|$, the two interactions are of similar strength resulting in observable interference effects, so called Coulomb-nuclear interference (CNI).

The TOTEM Collaboration has recently used the CNI to extract the value of the $\rho$ parameter, real-to-imaginary ratio of the forward amplitude, from elastic scattering differential cross-section at the collision energy $\sqrt s = 13\un{TeV}$ and interpreted the results as an argument in favour of the Odderon existence \cite{totem-13tev-rho}. This has revived also some theoretical interest in CNI; some recent publications are briefly discussed in the following paragraphs.

Petrov has studied CNI in an eikonal framework \cite{petrov2018,petrov2018-erratum}. Some of his results take a similar form to the formulae previously obtained by Cahn \cite{cahn82} and Kudr\'at-Lokaj\'i\v cek (KL) \cite{kl94}, but have one term less. Petrov argued that this is due to a wrong treatment of proton form factors in the work by Cahn. This hypothesis will be checked in this paper. Further details of the proposed mistakes in Cahn's derivations were given in Refs.~\cite{petrov2019,petrov2020}, in addition suggesting that the expansion in orders of the fine-structure constant, $\alpha$, was insufficiently truncated. Also this suggestion will be tested in the present paper.

Godizov has proposed that CNI effects may be negligible on amplitude level, since the Coulomb and nuclear eikonals have very little overlap \cite{godizov2019}. A similar statement has been made by Donnachie and Landshoff \cite{donnachie2019}. These proposals will be verified in this paper.

Khoze et al.~have re-confirmed the relevance of CNI amplitude effects and furthermore have evaluated the impact of inelastic intermediate states which are not taken into account in the traditional eikonal framework \cite{kmr2019}.

In this paper we focus on eikonal description of CNI, which is the common basis of works by Cahn, KL, Petrov and others. For a more complete historical review and other approaches see e.g.~Ref.~\cite{thesis}.

This paper follows an approach complementary to the aforementioned publications: instead of analytic manipulations, we present a numerical analysis starting with the fundamental assumption of the eikonal framework -- the additivity of eikonals (method first used in thesis \cite{thesis}). This approach allows to double-check the analytic derivations, some steps of which were found questionable even by the original authors, see e.g.~the comment above Eq.~(18) in Ref.~\cite{cahn82}.

Finally, the numerical approach used in this paper provides an explicit evaluation of the CNI to all orders of $\alpha$, to our knowledge, for the first time. Petrov has also provided a formula to all orders of $\alpha$ \cite{petrov2018} but, in our opinion, it is not well suited for numerical evaluation. Petrov has recently published another and more explicit CNI formula including all orders of $\alpha$ \cite{petrov2020}, but we have not had a chance to test its numerical properties yet.

After a preprint of this work has been made available \cite{preprint} Petrov has published a critical reaction \cite{petrov2020-2}. His critical comments are addressed in this revised version of the document.

The paper is organised as follows. In Section \ref{sec:eikonal} we briefly outline the essence of the eikonal framework. Section \ref{sec:results} will show predictions of different CNI formulae applied to nuclear amplitudes reflecting the TOTEM measurements at $\sqrt s = 8\un{TeV}$ \cite{totem-8tev-1km}. Section \ref{sec:technical} gives technical details of the numerical calculation. The paper is concluded with a summary in Section \ref{sec:summary}.

%----------------------------------------------------------------------------------------------------

\section{Eikonal calculation}
\label{sec:eikonal}

The CNI treatment in the eikonal framework can be sketched as follows.

The Coulomb amplitude in Born approximation, e.g.~from QED, is used as an input:
\begin{equation}
F^{\rm C}_{\rm Born}(t) = \pm {\alpha s\over t - \lambda^2} {\cal F}^2(t)\ ,
\label{eq:F C Born}
\end{equation}
where ${\cal F}$ stands for proton's form factor and the upper (lower) sign refers to proton-proton (proton-antiproton) scattering. The fictious photon mass, $\lambda$, is kept explicitly in the expression to act as an infrared regulator. The Coulomb eikonal can be obtained via \NEW{Fourier}-Bessel transform:
\begin{equation}
\delta^{\rm C}(b) = {1\over s} \int\limits_0^\infty \d q\, q\, J_0(bq)\, F^{\rm C}_{\rm Born}(-q^2)\ ,
\label{eq:de C}
\end{equation}
where $J_0$ is the zeroth order Bessel function of the first kind. In the special case with ${\cal F} \equiv 1$ (i.e.~point-like protons), the eikonal can be evaluated analytically \cite{cahn82}:
\begin{equation}
\delta^{\rm C}_{\rm asym}(b) = -\alpha K_0(\lambda b)\ ,
\label{eq:de C asym}
\end{equation}
where $K_0$ stands for the modified Bessel function of the second kind and zeroth order.

The nuclear amplitude in the impact-parameter space, $A^{\rm N}(b)$, can be obtained from the amplitude in the momentum space, $F^{\rm N}(t)$, with a \NEW{Fourier}-Bessel transform:
\begin{equation}
A^{\rm N}(b) = {1\over s} \int\limits_0^\infty \d q\, q\, J_0(bq)\, F^{\rm N}(-q^2)
\label{eq:A N}
\end{equation}
and the corresponding nuclear eikonal
\begin{equation}
\delta^{\rm N}(b) = {1\over 2\I} \log\left( 2\I A^{\rm N}(b) + 1\right)
\label{eq:de N}
\end{equation}

Following the assumed eikonal additivity, the total eikonal is obtained by summing the Coulomb and nuclear eikonals:
\begin{equation}
\delta^{\rm C+N}(b) = \delta^{\rm C}(b) + \delta^{\rm N}(b)\ .
\label{eq:de CN}
\end{equation}
The total amplitude, reflecting both Coulomb and nuclear interactions acting simultaneously, is given by the \NEW{inverse Fourier}-Bessel transform:
\begin{equation}
F^{\rm C+N}(t) = {s\over 2\I} \int
\limits
_0^\infty 
\d b\, b\,J_0(b\sqrt{-t})\,\left( \e^{2\I \delta^{\rm C+N}(b)} - 1 \right)\ .
\label{eq:F CN}
\end{equation}

Neglecting $\delta^{\rm N}$ in Eq.~(\ref{eq:de CN}), Eq.~(\ref{eq:F CN}) yields the complete Coulomb amplitude (i.e.~summation to all orders of $\alpha$). In the special case of ${\cal F} \equiv 1$, Cahn has found that the summation only affects the phase:
\begin{equation}
F^{\rm C}(t) = \pm {\alpha s\over t - \lambda^2} \e^{\I\alpha \eta(t)}\ ,\qquad \eta(t) = \log {\lambda^2\over -t}\ .
\label{eq:F C comp}
\end{equation}
Although this structure does not hold with a general form factor ${\cal F}$, Cahn used the following approximation for developing his CNI formula:
\begin{equation}
F^{\rm C}(t) \approx \pm {\alpha s\over t - \lambda^2} \e^{\I\alpha \eta(t)} {\cal F}^2  \ ,
\label{eq:F C comp ff}
\end{equation}
which is the subject of criticism by Petrov \cite{petrov2018}. The same approximation is found in the KL formula.

Differential cross-section is obtained from the corresponding amplitude by
\begin{equation}
{\d\sigma\over\d t} = {\pi(\hslash c)^2\over s p^2} |F|^2\ .
\label{eq:cs}
\end{equation}

%----------------------------------------------------------------------------------------------------

\section{Results}
\label{sec:results}

\begin{figure*}[h]
\begin{center}
\includegraphics{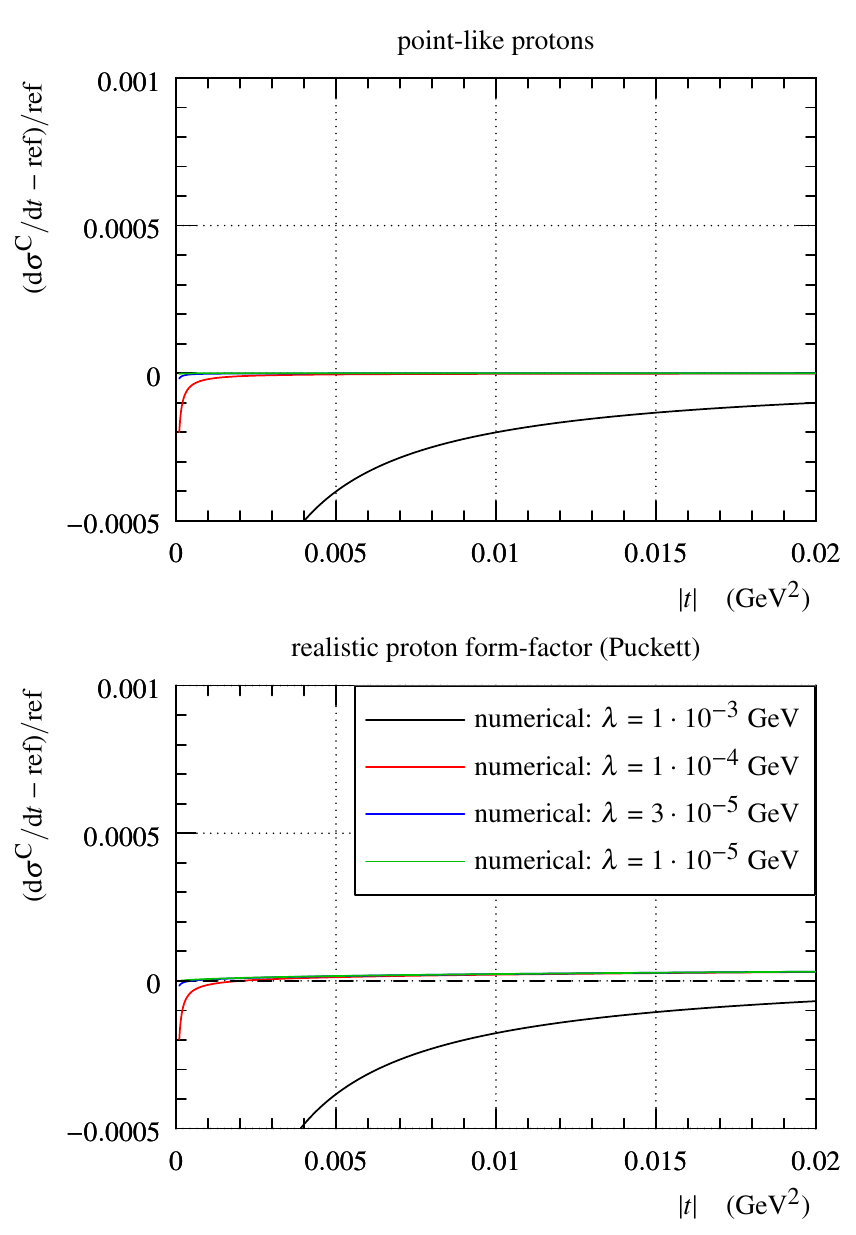}
\caption{Complete Coulomb cross-section -- relative difference between numerical calculation (Eq.~(\ref{eq:F CN}) with $\delta^{\rm N} \equiv 0$) wrt.~Born-level input, Eq.~(\ref{eq:F C Born}) with $\lambda = 0$. The different colours represent different choices of $\lambda$. {\it Top}: for point-like charges, ${\cal F} \equiv 1$, {\it bottom}: with a realistic proton form factor.}
\label{f:sig C}
\end{center}
\end{figure*}

\begin{figure*}[h]
\begin{center}
\includegraphics{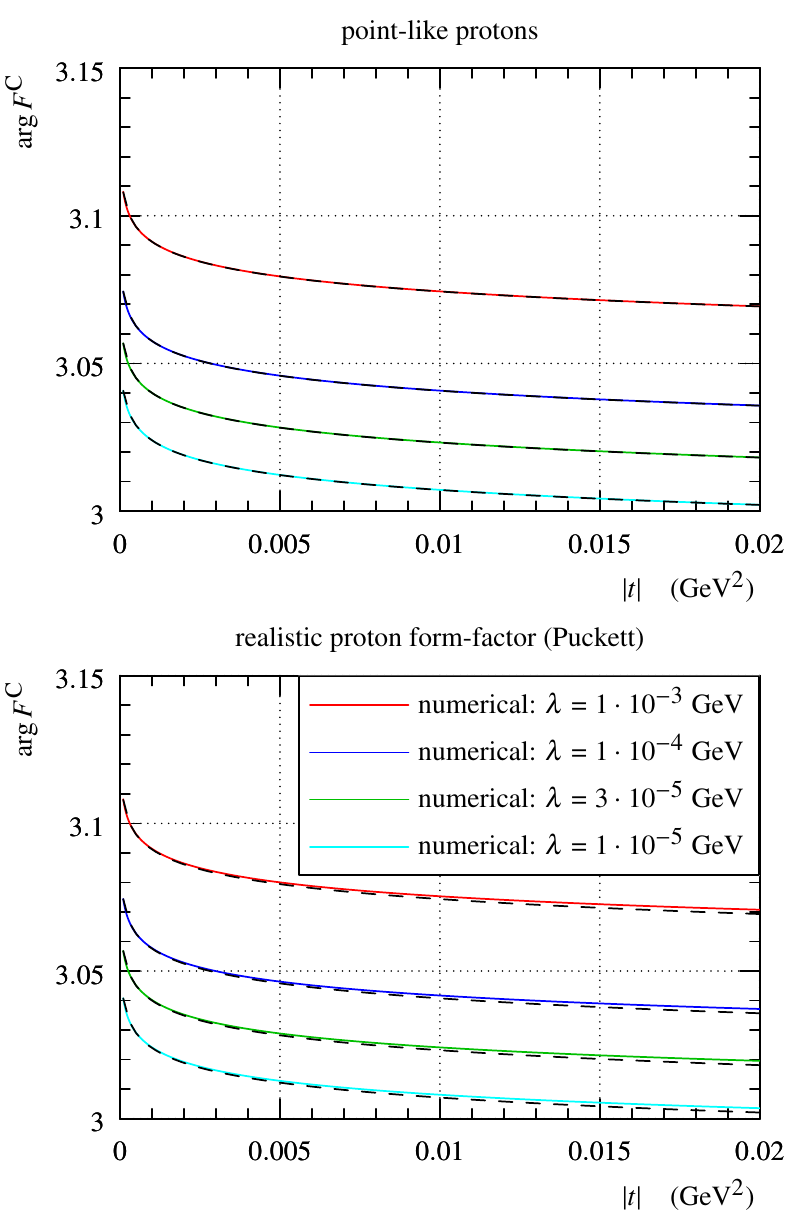}
\caption{Phase of the complete Coulomb amplitude. Coloured lines come from numerical calculation (with different choices of $\lambda$). The black dashed curves correspond to $\pi + \alpha\eta(t)$, the phase of the amplitude in Eqs.~(\ref{eq:F C comp}) and (\ref{eq:F C comp ff}). {\it Top}: for point-like charges, ${\cal F} \equiv 1$, {\it bottom}: with a realistic proton form factor.}
\label{f:arg F C}
\end{center}
\end{figure*}

In this section, predictions from several CNI formulae will be compared:
\begin{itemize}
\item ``numerical'': numerical evaluation of Eq.~(\ref{eq:F CN}),
\item ``Cahn'': Eq.~(30) in Ref.~\cite{cahn82},
\item ``KL'': Eq.~(26) in Ref.~\cite{kl94},
\item ``Petrov'': Eq.~(17) in Ref.~\cite{petrov2018} (taking into account the erratum \cite{petrov2018-erratum})
\item ``SWY'': Eq.~(26) in Ref.~\cite{wy68},
\item ``trivial'': plain sum of the Coulomb and nuclear amplitude, as suggested e.g.~in Ref.~\cite{godizov2019}.
\end{itemize}
For completness: Eq.~(13) in Ref.~\cite{petrov2018} (to all orders of $\alpha$) is not considered here -- not for lack of interest, but for difficulties in constructing a corresponding numerical-integration computer program. We believe that it is worth a forthcoming study, along with considering another represation proposed by Petrov, e.g.~Eq.~(13) in Ref.~\cite{petrov2020-2}.

To test the numerical calculation one needs to assume a certain nuclear amplitude, $F^{\rm N}(t)$. This unavoidably introduces some model-dependence in our results. In order to focus on physics-relevant models, we will use the two nuclear amplitudes published by the TOTEM Collaboration in an analysis of $\sqrt s = 8\un{TeV}$ proton-proton data \cite{totem-8tev-1km} (Table 5). While the differential cross-section measurement puts strict constraints on the amplitude modulus, the phase remains almost arbitrary. Consequently, two extreme/alternative options will be tested: ``central'' with nuclear phase constant in $t$ and ``peripheral'' with nuclear phase rapidly varying in $t$. The labels have been chosen to reflect the different impact-parameter behaviour: the ``central'' model yields a profile function peaking at smaller impact-parameter value wrt.~the ``peripheral'' model.

The proton form-factor will be modelled according to Puckett et al.~\cite{puckett}.

In the numerical calculation, the $\lambda$ regulator cannot be strictly set to zero, but instead it can be chosen small enough not to have any significant impact on the results in the $b$ and $t$ ranges of interest. This is illustrated for example in Figure \ref{f:sig C}: results for different values of $\lambda$ are shown in different colours. As $\lambda$ gets smaller, the difference between results diminishes. In particular, there is almost no visible difference between $\lambda = 3\cdot 10^{-5}$ (blue) and $10^{-5}\un{GeV}$ (green). This indicates that the former value of $\lambda$ is small enough (for our $|t|$ range) and will be often used as a reference for comparisons.

Figure \ref{f:sig C} compares the complete (i.e.~to all orders of $\alpha$) Coulomb cross-section from the numerical calculation (colours) to the input Born-level expression (black dashed). The top plot, corresponding to point-like protons, shows a perfect agreement between the numerical calculation (for sufficiently small $\lambda$) and the Born curve, as expected from Eq.~(\ref{eq:F C comp}). The bottom plot, corresponding to a realistic proton form factor, shows small relative deviations, ${\cal O}(10^{-4})$.

Figure \ref{f:arg F C} shows the phase of the complete Coulomb amplitude which depends on the choice of $\lambda$ (different colours). The top plot, for point-like protons, indicates a perfect agreement with the $\eta(t)$ calculation by Cahn (black dashed). The bottom plot, for a realistic form factor, shows small deviations, ${\cal O}(10^{-3})$.

\begin{figure*}[h]
\begin{center}
\includegraphics{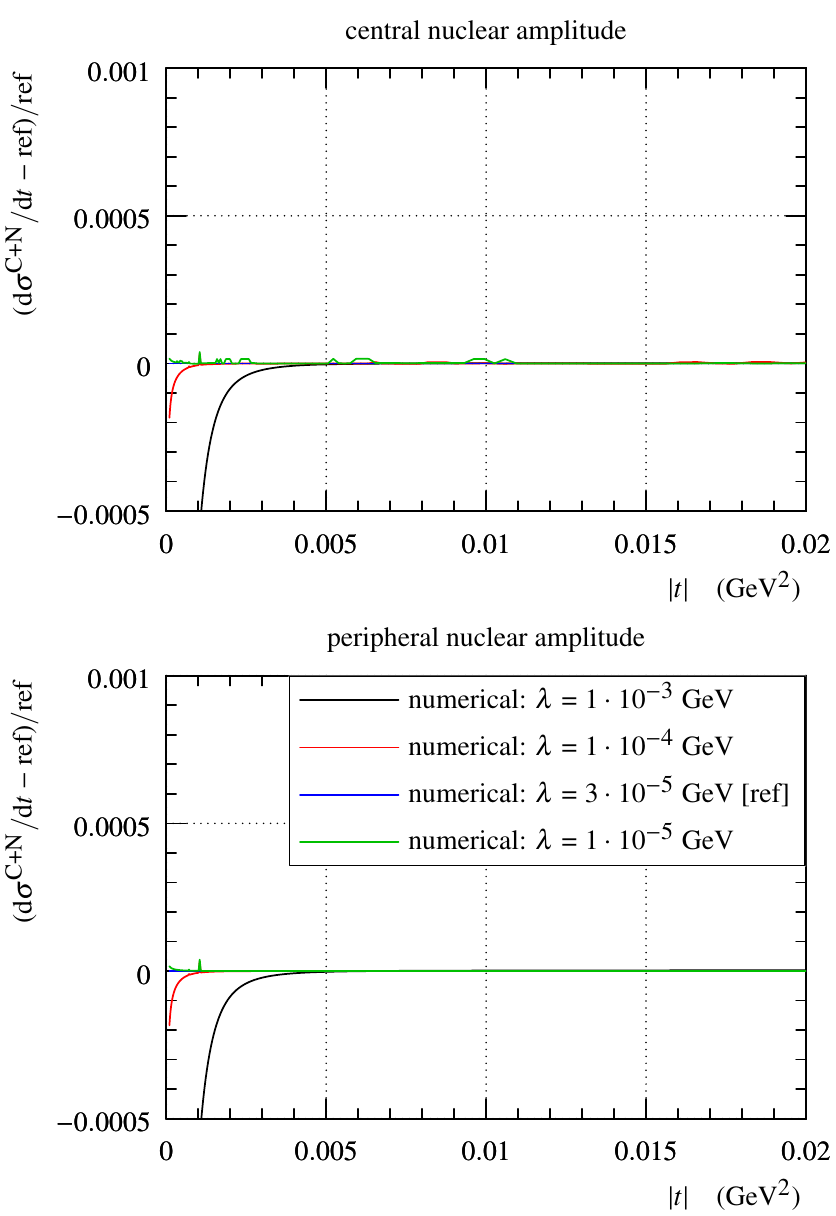}
\caption{Full Coulomb+nuclear cross-section as obtained from the numerical calculation, Eq.~(\ref{eq:F CN}), for different values of $\lambda$ (colours). The green curve suffers from little numerical instabilities (negligible compared to typical experimental uncertainties). {\it Top}: for central nuclear amplitude, {\it bottom}: for peripheral nuclear amplitude.}
\label{f:sig CN}
\end{center}
\end{figure*}

Figure \ref{f:sig CN} compares the total (Coulomb + nuclear) cross-section from the numerical calculation for several choices of $\lambda$. Like in Figure \ref{f:sig C}, the smaller $\lambda$, the smaller difference in predictions. When $\lambda \lesssim 3\cdot 10^{-5}\un{GeV}$, almost no visible difference is present. This has been further verified for $\lambda$ values down to $3\cdot10^{-6}\un{GeV}$ and agrees with the expectation from Eq.~(\ref{eq:F C Born}): introducing $\lambda$ makes negligible effect whenever $\lambda^2 \ll |t|$. In conclusion, we believe that we can choose $\lambda$ sufficiently small such that the numerical calculation gives predictions comparable (on our $|t|$ range starting at $10^{-4}\un{GeV^2}$) with the $\lambda\to 0$ limit. We find misleading the comment below Eq.~(1) in Ref.~\cite{petrov2020-2} that a single value of $\lambda$ is used. In contrary, a series of $\lambda$ values is systematically considered and the corresponding results are analysed. Only the fact the results evolve with $\lambda$ as expected allows us to make the presented interpretation.

\begin{figure*}[h]
\begin{center}
\includegraphics{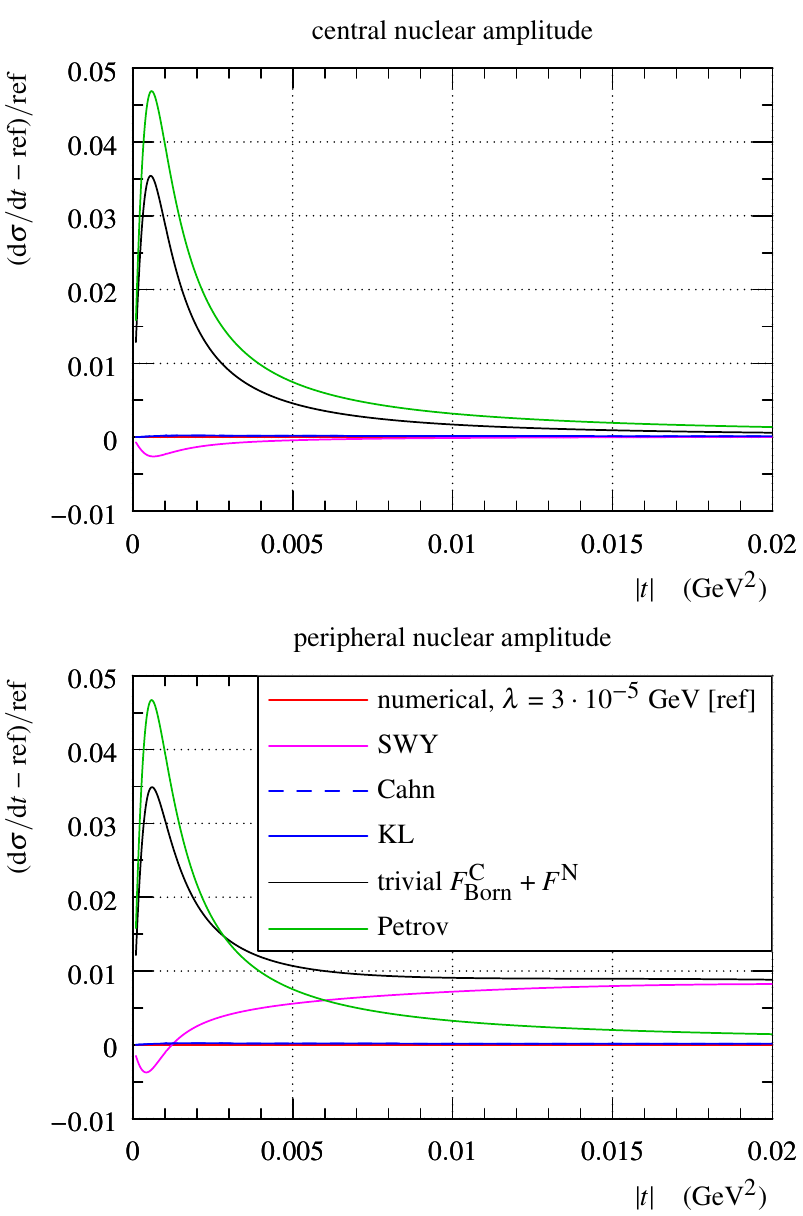}
\caption{Relative difference between various CNI formulae and the reference from the numerical calculation (red). {\it Top}: for central nuclear amplitude, {\it bottom}: for peripheral nuclear amplitude.}
\label{f:sig form}
\end{center}
\end{figure*}

Figure \ref{f:sig form} compares predictions from several CNI formulae to the reference from the numerical calculation (red, to all orders of $\alpha$). For both central (top) and peripheral (bottom) cases, the predictions by Cahn and KL are almost identical and they overlap with the numerical-calculation reference -- the relative difference is ${\cal O}(10^{-4})$. The trivial sum of the Coulomb and nuclear amplitudes can deviate up to about $3.5\un{\%}$. The formula by Petrov (missing one term wrt.~Cahn/KL) can deviate by almost $5\un{\%}$. The SWY formula provides relatively good description in the ``central'' case (relative deviations ${\cal O}(10^{-3})$) and somewhat worse description in the ``peripheral'' case (deviations up to about $1\un{\%}$). This is not surprising since the SWY formula assumes a slow nuclear phase variation.

\NEW{%
Let us emphasize that the numerical calculation presented in this article, the work by Cahn \cite{cahn82}, the work by Kundr\'at and Lokaj\'i\v cek \cite{kl94} and the work by Petrov \cite{petrov2018} are all based on the identical set of assumptions and expressions: the additivity of eikonals\footnote{\NEW{%
Our Eq.~(\ref{eq:de CN}) is equivalent to Eq.~(12) in Ref.~\cite{cahn82} (implicitely), to Eq.~(7) in Ref.~\cite{kl94} and to Eq.~(3) in Ref.~\cite{petrov2018}.
}} and the expression for the Coulomb eikonal\footnote{\NEW{%
Our Eq.~(\ref{eq:de C}) is equivalent to Eq.~(10) in Ref.~\cite{cahn82}, to Eq.~(12) in Ref.~\cite{kl94} and to Eq.~(12) in Ref.~\cite{petrov2018}.
}}. Therefore the differences reported in the previous paragraph can only be due to approximations (analytic or numerical) or mistakes in the corresponding works.
}

We disagree with the statement made below Eq.~(17) in Ref.~\cite{petrov2020-2}: if a formula shall hold generally and if we find (with whatever method) even a single example where it fails, we believe that this generally undermines the validity of the formula. In this way we interpret the deviations reported \NEW{above}. Furthermore, let us stress that the examples used in this document are not ``random'' but with sound physics motivation. We do realise that numerical calculations come with limited accuracy -- for that reason multiple checks and validations are presented throughout this article.

%----------------------------------------------------------------------------------------------------

\section{Technical details}
\label{sec:technical}

The numerical integration in Eqs.~(\ref{eq:de C}), (\ref{eq:A N}) and (\ref{eq:F CN}) is performed with the help of the GSL library \cite{gsl}, in particular using adaptive integration based on 61-point Gauss-Kronrod rules.

For the numerical integration one needs to set reasonable boundaries. In the case of Eq.~(\ref{eq:F CN}) a reasonable upper limit, $b_{\rm max}$, can be deduced by analysing the expression in the parentheses, in the lowest order being $2\I\delta^{\rm C+N}$. Since the nuclear interaction is expected to be short-ranged, $\delta^{\rm N}$ shall vanish at large $b$ and thus $\delta^{\rm C+N} \approx \delta^{\rm C}$. Furthermore, the effect of Coulomb form-factors is expected to be concentrated at low $b$, therefore at large $b$ one may safely approximate $\delta^{\rm C+N} \approx \delta^{\rm C}_{\rm asym}$. These assumptions have been explicitly tested for the choices of nuclear amplitudes and Coulomb form-factors used in this article. Since $\delta^{\rm C}_{\rm asym} \propto K_0(\lambda b)$ and since $K_0$ is a monotonously falling function, one may truncate the integration once the $K_0(\lambda b)$ function becomes sufficiently small, i.e.~when $\lambda b$ exceeds a certain threshold. Consequently, we adopted $b_{\rm max} = c / \lambda$, where $c = 10$ was found appropriate by numerical tests -- variation of $c$ between $5$ and $50$ leads to negligible changes in the results. Finally, we would like to stress that this paragraph is about setting numerical integration bounds for Eq.~(\ref{eq:F CN}). For different integrations, different rules should be used -- in that sense we do not think that the criticism below Eq.~(18) in Ref.~\cite{petrov2020-2} is applicable here.

In the case of Eq.~(\ref{eq:de C}), the upper limit was set to $q_{\rm max} = 10^{max(3, 3 - log_{10}(b))} \un{GeV}$. This rule was found with numerical tests, there is negligible variation of the results when the parameters and varied around the quoted values. The rule works both with and without including form-factors. The reduction of $q_{\rm max}$ with $b$ can be justified by the fact that the amplitude of $J_0(bq)$ oscillations decreases with increasing $bq$.

The implementation of the analytic interference formulae (Cahn, KL, Petrov and SWY) is based on the Elegent software package \cite{elegent}.

Several optimisations are used in the numerical evaluation. First, the asymptotic expression $\delta^{\rm C}_{\rm asym}$ is used instead of the integral in Eq.~(\ref{eq:de C}) for $b > 20\un{GeV^{-1}}$. It has been checked that the relative error of this simplification is smaller than $10^{-4}$. Then, Eq.~(\ref{eq:F CN}) is recast such that the expression in the parentheses is reduced by $2\I\delta^{\rm C}$ which is compensated by adding the Coulomb Born amplitude, Eq.~(\ref{eq:F C Born}), outside the integral. This algebraic transformation improves the convergence of the numerical integration.

The full calculation code in C++ is available in a public GitHub repository \cite{code}.

%----------------------------------------------------------------------------------------------------

\section{Summary and conclusions}
\label{sec:summary}

It has been verified with a realistic proton form-factor that Cahn's approximation of the complete Coulomb amplitude, Eq.~(\ref{eq:F C comp ff}), is inexact as argued by Petrov \cite{petrov2018}. However, the deviation is rather small: ${\cal O}(10^{-3})$ for phase and ${\cal O}(10^{-4})$ for the relative deviation in modulus. Such deviations are likely to be undetectable with the current experimental possibilities.

A numerical eikonal calculation of CNI effects, based directly on the eikonal additivity and carried out to all orders of $\alpha$ has been presented, likely for the first time.

The new CNI formula proposed by Petrov \cite{petrov2018} (with one term missing wrt.~the formula by Cahn/KL) has been compared with the numerical calculation and found to deviate up to almost $5\un{\%}$.

A plain sum of the Coulomb and nuclear amplitudes, compared to the eikonal numerical calculation, leads to deviations up to $3.5\un{\%}$. \NEW{We consider this observation as an indication that the proposal is oversimplified}.

The SWY formula reproduces the numerical eikonal calculation well for the ``central'' nuclear amplitude. In the ``peripheral'' case, the deviations are up to $1\un{\%}$.

The best reproduction of the numerical eikonal calculation has been found by the Cahn/KL formulae, the relative deviations are ${\cal O}(10^{-4})$. This indicates that Cahn's inexact approximation of the complete Coulomb amplitude and the early truncation of the series in powers of $\alpha$ (as pointed out by Petrov \cite{petrov2019,petrov2020}) do not have any detrimental effect that could be currently experimentally observed. This leads us to the conclusion that the formulae by Cahn/KL are currently the ``best on the market''.

\NEW{Since the numerical calculation presented in this article and the interference formulae by Cahn, Kundr\'at-Lokaj\'i\v cek and Petrov are all based on the same premisses, one should expect identical results (within the uncertainty due to the analytic/numerical approximations applied). We find that the numerical calculation agrees well with the Cahn/KL formulae, but all of them differ significantly from the one by Petrov. Although this is shown with only two concrete examples (still physically very relevant), we interpret this as a general failure of the formula proposed by Petrov.
}

One may argue that taking the eikonal calculation as reference is a biased choice, since the eikonal framework is an approximation on its own and it cannot naturally include some of the known effects (further discussion can be found e.g.~in Refs.~\cite{thesis,petrov2018,kmr2019}). Possibly one of the effects most difficult to evaluate -- the influence of the inelastic intermediate states -- has recently been estimated by Khoze et al.~\cite{kmr2019}, finding that the effect would not be observable with the current experimental accuracy.

Overall, we find that TOTEM has chosen a reasonable model of CNI effects to extract the $\rho$ parameter \cite{totem-13tev-rho}.

%----------------------------------------------------------------------------------------------------

\section{Acknowledgements}

The author is grateful for stimulating discussions with A.~Godizov, V.~Khoze and collaborators, V.~Kundr\'at and V.~Petrov. The author also wishes to thank several of them for valuable suggestions how to improve this manuscript.

%----------------------------------------------------------------------------------------------------

\def\journal#1#2#3#4{
	#1 #2 (#3) #4
}

\end{document}